\documentclass[10pt,letterpaper]{article}
\usepackage{opex3}

\usepackage{here}
\usepackage{graphicx}
\graphicspath{{pict/}{}}

\newcommand\pictc[5]{\begin{figure}[t]
            \centerline{\vspace{0mm}
\includegraphics[width=#1\columnwidth,height=0.7\textheight,keepaspectratio]{#3}}
            \protect\caption{\protect\label{fig:#4} #5}
                    \end{figure}            }

\newcommand\pict[4][0.57]{\pictc{#1}{!tb}{#2}{#3}{#4}}
\newcommand\rpict[1]{\ref{fig:#1}}

\newcommand\leqt[1]{\protect\label{eq:#1}}
\newcommand\reqtn[1]{\ref{eq:#1}}
\newcommand\reqt[1]{(\reqtn{#1})}

\newcounter{Fig}

\begin{document}

\begin{sloppy}

\title{Surface multi-gap vector solitons}

\author{Ivan L. Garanovich, Andrey A. Sukhorukov, and Yuri S. Kivshar}
\address{\mbox{Nonlinear~Physics~Centre~and~Centre~for Ultra-high bandwidth~Devices~for~Optical~Systems~(CUDOS)},
Research~School~of~Physical~Sciences~and~Engineering, Australian~National~University,
Canberra, ACT 0200, Australia}
\email{ilg124@rsphysse.anu.edu.au}

\author{Mario Molina}
\address{Departamento de F\'{\i}sica, Facultad de Ciencias, Universidad
de Chile, Santiago, Chile}

\email{mmolina@uchile.cl}

\begin{abstract}
We analyze nonlinear collective effects near surfaces of
semi-infinite periodic systems with multi-gap transmission spectra
and introduce a novel concept of {\em multi-gap surface solitons} as
mutually trapped surface states with the components associated with
different spectral gaps. We find numerically discrete surface modes
in semi-infinite binary waveguide arrays which can support
simultaneously two types of discrete solitons, and analyze different
multi-gap states including the soliton-induced waveguides with the
guided modes from different gaps and composite vector solitons.
\end{abstract}

\ocis{\footnotesize (190.4420) Nonlinear optics, transverse effects in;
      (190.5940) Self-action effects
     }

\section{Introduction}

Interfaces separating different physical media can support a special
class of transversally localized waves known as {\em surface waves}.
Linear surface waves have been studied extensively in many branches
of physics~\cite{Davidson:1996:SurfaceStates}, and the structure of
surface states in periodic systems is associated with the specific
properties of the corresponding Bloch waves. Such linear surface
waves with {\em staggered} profiles are often referred to as {\em
Tamm states}~\cite{Tamm:1932-849:ZPhys}, first identified as
localized electronic states at the edge of a truncated periodic
potential, and then found in other systems, e.g. for an interface
separating periodic and homogeneous dielectric optical
media~\cite{Yeh:1977-423:JOS, Yeh:1978-104:APL}.

{\em Nonlinear surface waves} have been studied most extensively in
optics where both TE and TM surface modes were predicted and
analyzed for the interfaces separating two different homogeneous
nonlinear dielectric media~\cite{Tomlinson:1980-323:OL,
Boardman:1991-73:NonlinearSurface,
Mihalache:1989-229:ProgressOptics}. In addition, nonlinear effects
are known to stabilize surface modes in discrete systems generating
different types of states localized at and near the
surface~\cite{Kivshar:1998-125:PD}. Self-trapping of light near the
boundary of a self-focusing photonic lattice has recently been
predicted theoretically~\cite{Makris:2005-2466:OL} and demonstrated
in experiment~\cite{Suntsov:2006-63901:PRL} through the formation of
{\em discrete surface solitons} at the edge of a waveguide array.
Such {\em unstaggered} discrete surface modes can be treated as
discrete optical solitons~\cite{Kivshar:2003:OpticalSolitons}
trapped at the edge of a waveguide array when the beam power exceeds
a certain critical value associated with a strong repulsive
surface energy~\cite{Molina:2006-discrete:OL}. Staggered surface gap
solitons in defocusing semi-infinite periodic media have also been
introduced theoretically~\cite{Kartashov:2006-73901:PRL} and
recently observed experimentally~\cite{Rosberg:2006-xxx:ProcCLEO},
and they provide a direct nonlinear generalization of the familiar
electronic surface Tamm states.

So far, unlike the only case of vector discrete nonlinear surface
waves~\cite{Hudock:2005-7720:OE}, discrete surface solitons have
been described by scalar fields, and they were associated either
with the semi-infinite total-internal reflection optical gap, such
as discrete surface solitons~\cite{Makris:2005-2466:OL}, or with the
Bragg-reflection photonic gap, such as surface gap
solitons~\cite{Kartashov:2006-73901:PRL}, being treated completely
independently. However, as was already shown for infinite
nonlinear periodic and discrete systems~\cite{Cohen:2003-113901:PRL,
Sukhorukov:2003-113902:PRL, Cohen:2005-500:NAT, Rosberg:2005-5369:OE}, nonlinear collective effects in photonic systems with multi-gap transmission spectra may lead to mutual cross-band focusing effects and can
support {\em multi-gap solitons} as composite modes with the
components associated with different spectral gaps.

In this paper, we introduce the concept of {\em multi-gap surface
waves} as composite states consisting of mutually trapped components
from different gaps localized at the surface. Such composite surface
states can be created either by a surface soliton that traps linear
guided modes from other spectral gaps, or as vector surface solitons
with the major components from different gaps. To the best of our
knowledge, such states have never been mentioned in any field of
physics, and they become possible due to interaction of multi-gap
discrete solitons~\cite{Sukhorukov:2003-113902:PRL} with the surface. As a specific
example allowing us to demonstrate the basic concept, we consider
here the discrete surface modes in semi-infinite binary waveguide
arrays, earlier introduced theoretically~\cite{Sukhorukov:2002-2112:OL, Sukhorukov:2003-2345:OL} and
then studied experimentally~\cite{Morandotti:2004-2890:OL}, and find numerically
different classes of multi-gap surface states also discussing their
existence and stability.

\pict[0.71]{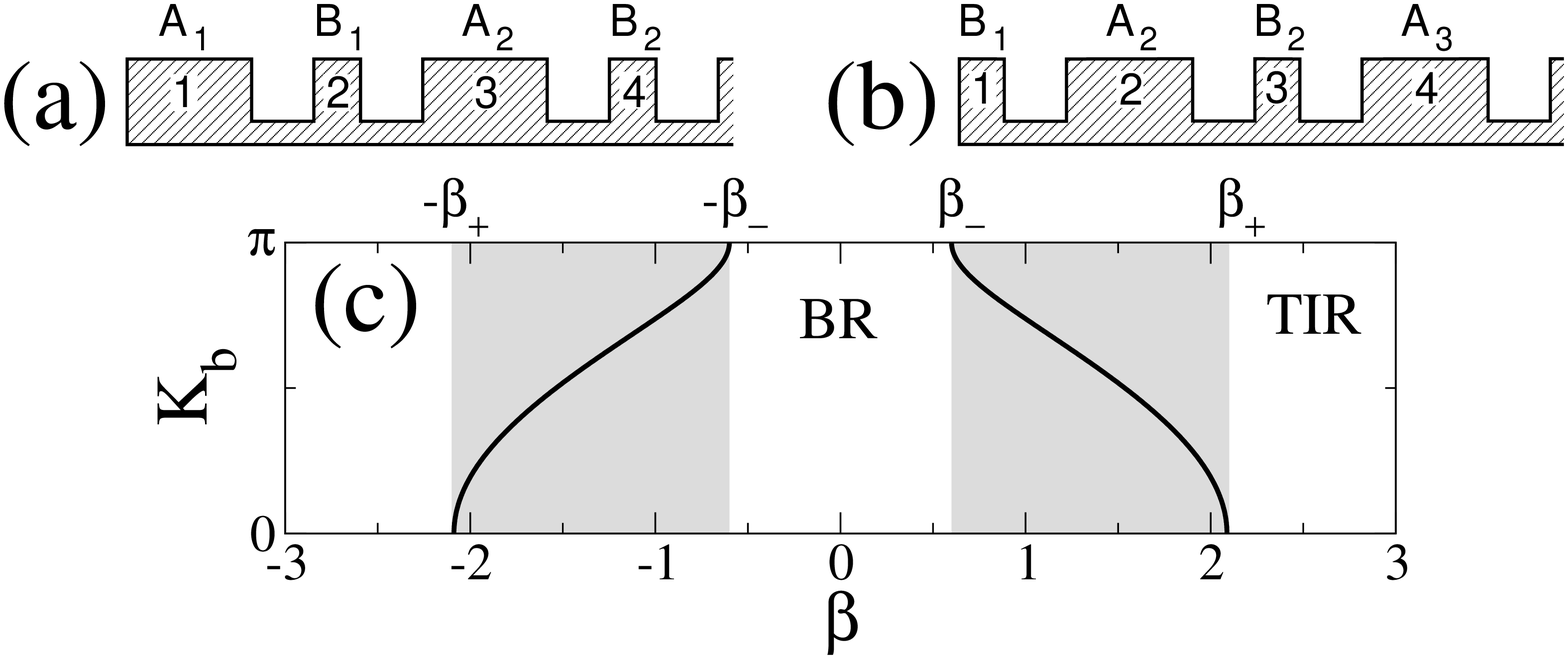}{binaryDispersion}{ (a,b) Two types of a
semi-infinite binary array truncated at wide or narrow waveguides.
(c) Dispersion diagram of the binary array with the bands shaded
and gaps marked. Detuning between the A and B-type waveguides is
$\rho=0.6$.}

\section{Surface discrete solitons in binary arrays}

We consider a binary array composed of two types (A and B, wide and
narrow) of separated optical
waveguides~\cite{Sukhorukov:2002-2112:OL, Sukhorukov:2003-2345:OL},
that can be fabricated by etching waveguides on top of a AlGaAs
substrate~\cite{Morandotti:2004-2890:OL}. To study the surface
effects, we assume that the array is truncated at either waveguide,
as shown in Figs.~\rpict{binaryDispersion}(a,b). We analyze
interactions between several mutually incoherent components with the
electric field envelopes $E^{(j)}(x,z)$. Then, we employ the
tight-binding approximation and present the total field ${\bf E}$ as
a superposition of the guided modes supported by individual guides,
${{\bf E}(x,z)} = \sum_n [{{\bf A}_n}(z) {\bf \psi}_{A_n}(x) + {{\bf B}_n}(z)
{\bf \psi}_{B_n}(x)] \exp(i k z)$, where we introduce the vector
notations ${\bf F} = (F^{(1)},F^{(2)},\ldots)$. Here ${\bf
\psi}_{A_n}(x)$ and ${\bf \psi}_{B_n}(x)$ are the mode profiles of the
individual waveguides, $k$ is the average propagation constant of A
and B modes, and $A_n^{(j)}$ and $B_n^{(j)}$ are the mode amplitudes
for the field component number $j$. Finally, we derive a system of
coupled discrete equations~\cite{Sukhorukov:2003-113902:PRL} for the
normalized amplitudes ${{\bf a}_n}$ and ${{\bf b}_n}$,
\begin{equation} \leqt{dnls}
   \begin{array}{l} {\displaystyle
      i \frac{d {\bf a}_n}{d z}
      + \rho {\bf a}_n
      + {\bf b}_{n-1} + {\bf b}_n
      + ||{\bf a}_n||^2 {\bf a}_n
      = 0 ,
   } \\*[9pt] {\displaystyle
      i \frac{d {\bf b}_n}{d z}
      - \rho {\bf b}_n
      + {\bf a}_{n} + {\bf a}_{n+1}
      + ||{\bf b}_n||^2 {\bf b}_n
      = 0,
   } \end{array}
\end{equation}
where ${\bf b}_0 \equiv 0$ or ${\bf a}_{1} \equiv 0$ for arrays terminated at a wide or narrow waveguide, respectively.
We consider the case of a positive Kerr-type medium response
proportional to the total field intensity $||{\bf \{a,b\}}_n||^2 =
\sum_j |{\{a,b\}}_n^{(j)}|^2$, neglecting the small differences in
the effective nonlinear coefficients at the A and B sites; $\rho$ is
proportional to the detuning between the propagation constants of
the A and B-type guided modes.

According to Eqs.~\reqt{dnls}, the linear Bloch-wave dispersion is
defined as $K_b = \cos^{-1}( -\eta / 2)$, where $\eta = 2 + \rho^2 -
\beta^2$. The transmission bands correspond to real $K_b$, and they
appear when $\beta_- \le |\beta| \le \beta_+$, where $\beta_- =
|\rho|$ and $\beta_+ = (\rho^2 + 4)^{1/2}$. A characteristic
dispersion relation and the corresponding band-gap structure are
presented in Fig.~\rpict{binaryDispersion}(c). The upper gap at
$\beta>\beta_+$ is due to the effect of  total internal reflection
(TIR gap), whereas additional BR gap appears for $|\beta|<\beta_-$
due do the resonant Bragg reflection.

First, we study scalar discrete surface solitons in the truncated
binary arrays and look for spatially localized solutions 
of Eqs.~\reqt{dnls} in the form $(a_n, b_n) = (u_{2 n-1-n_0}, u_{2
n-n_0}) \exp(i \beta z)$, $n=1,2\ldots$, where $\beta$ is the
propagation constant, $n_0=0,1$ for the structure termination at the
A or B site, respectively, and the function $u_n$ describes the
soliton profiles. Discrete solitons can appear in the TIR gap, and
gap solitons can form inside the BR gap.

\pict{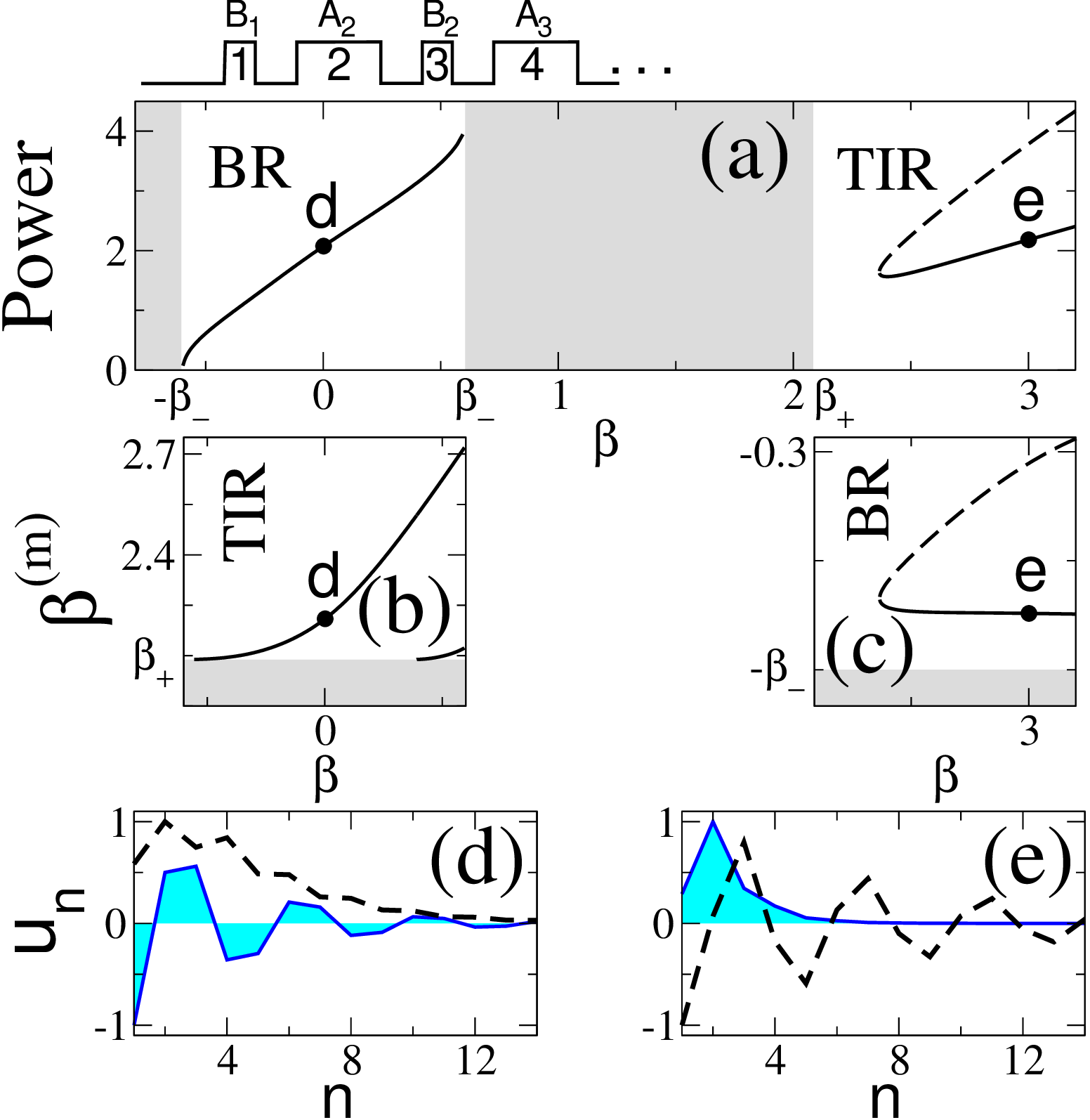}{solitonsModesNarrow}{
(a) Families of the discrete surface solitons in the BR and
TIR gaps, and (b,c) propagation constants of the localized linear modes from the other gaps 
guided by these solitons, in the binary array truncated at the narrow waveguide.
Dashed curves in (a-c) mark unstable soliton branches.
(d,e)~Examples of two types of the discrete surface solitons (solid) and their guided
linear modes localized in the other gaps (dashed), corresponding to the marked points "d" and "e" in (a-c).
Detuning parameter of the binary array is $\rho=0.6$.}

\pict{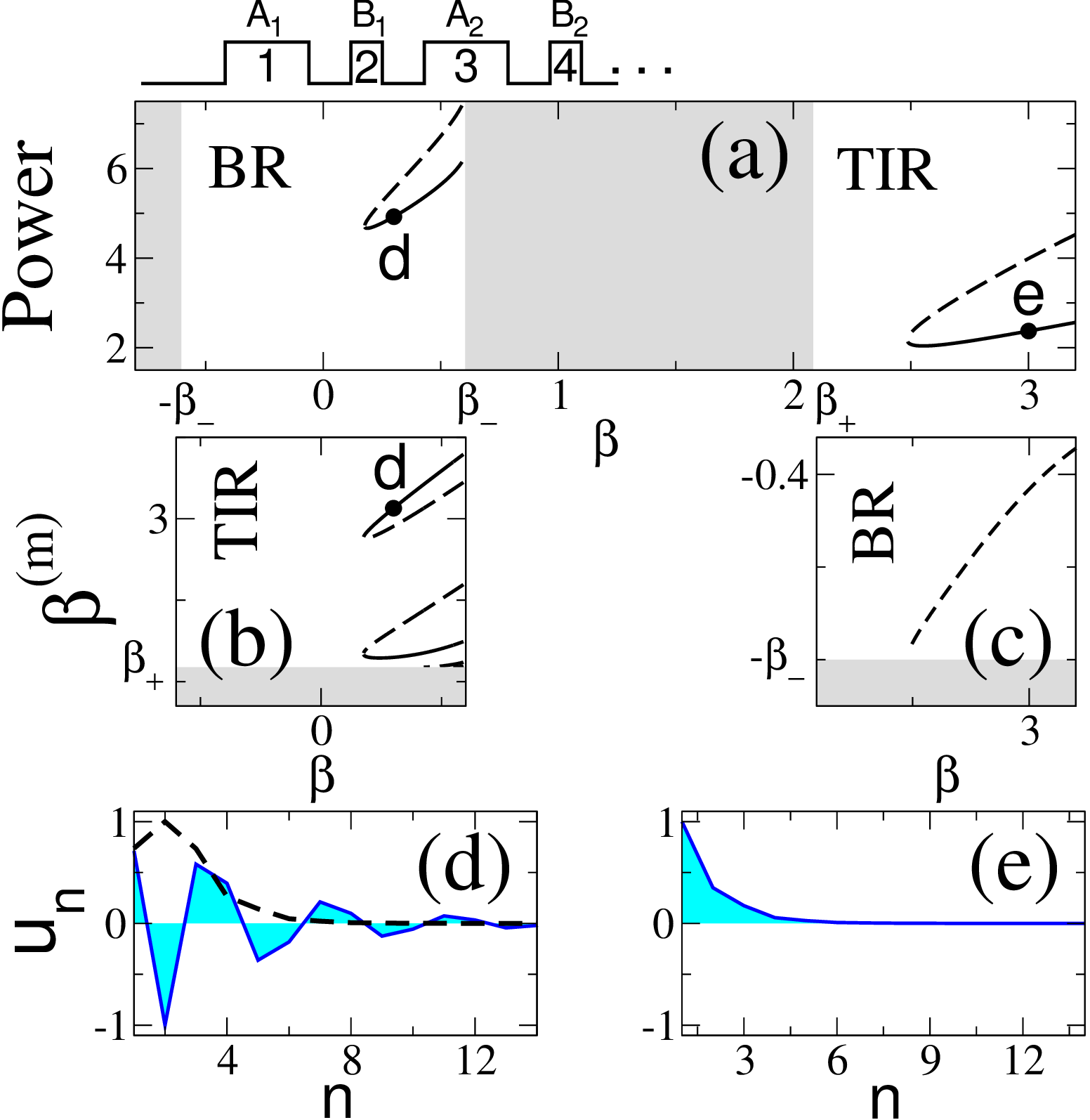}{solitonsModesWide}{ Same as in
Fig.~\rpict{solitonsModesNarrow}, but for the binary array truncated
at the wide waveguide.}

In Figs.~\rpict{solitonsModesNarrow}(a) and
\rpict{solitonsModesWide}(a) we summarize our findings and show the
results for the families of the nonlinear surface states for the two
different cases. In the first case, when the edge waveguide is
narrow, the nonlinear surface modes start to appear in the BR gap (as the
nonlinear Tamm states) at low powers, whereas the existence of the
surface modes in the TIR gap requires the beam power to exceed some
threshold value. In the second case, when the array is truncated at
the wide waveguide, the existence of nonlinear surface modes in both
the gaps requires similarly to exceed a threshold power. We 
employ the beam propagation method to investigate the soliton stability.
In Figs.~\rpict{solitonsModesNarrow}(a) and
\rpict{solitonsModesWide}(a), dashed branches of the curves indicate
unstable surface modes. Whereas oscillatory instabilities~\cite{Pelinovsky:2004-36618:PRE} may arise
for the solid branches, we have
verified that the solitons corresponding to marked points on the
solid curves demonstrate stable propagation for more than 100
coupling lengths even with input perturbations of 1\%.

\section{Waveguides induced by surface solitons}

The mutual trapping of the modes localized in different gaps and the
physics of surface multi-gap vector solitons can be understood in
terms of the soliton-induced waveguides. We search for the linear guided modes 
supported by a scalar soliton in other gaps, 
and consider two types of surface solitons and two different
waveguide truncations. In Figs.~\rpict{solitonsModesNarrow}(b,c)
and~\rpict{solitonsModesWide}(b,c), we plot the eigenvalues of the
linear guided modes supported by the BR~(b) and TIR~(c) surface
solitons, respectively. In Figs.~\rpict{solitonsModesNarrow}(d,e)
and~\rpict{solitonsModesWide}(d,e) we show several characteristic
examples of the BR and TIR surface solitons together with linear
surface modes they guide in the other gap. We find that surface
solitons may be able to guide linear surface modes,  and the number of
guided modes supported by a soliton-induced surface waveguide
depends on the lattice termination and the soliton distance from the
edge, indicating the possibility of the effective engineering of the interband
interactions. For example, in contrast to the case of infinite waveguide arrays where linear modes are always present~\cite{Cohen:2003-113901:PRL, Sukhorukov:2003-113902:PRL, Pelinovsky:2004-36618:PRE}, in the case of the structure termination on the A-type (narrow) waveguide stable fundamental TIR solitons do not support BR modes [note the absence of a solid curve in Fig.~\rpict{solitonsModesWide}(c)].

\pict[0.61]{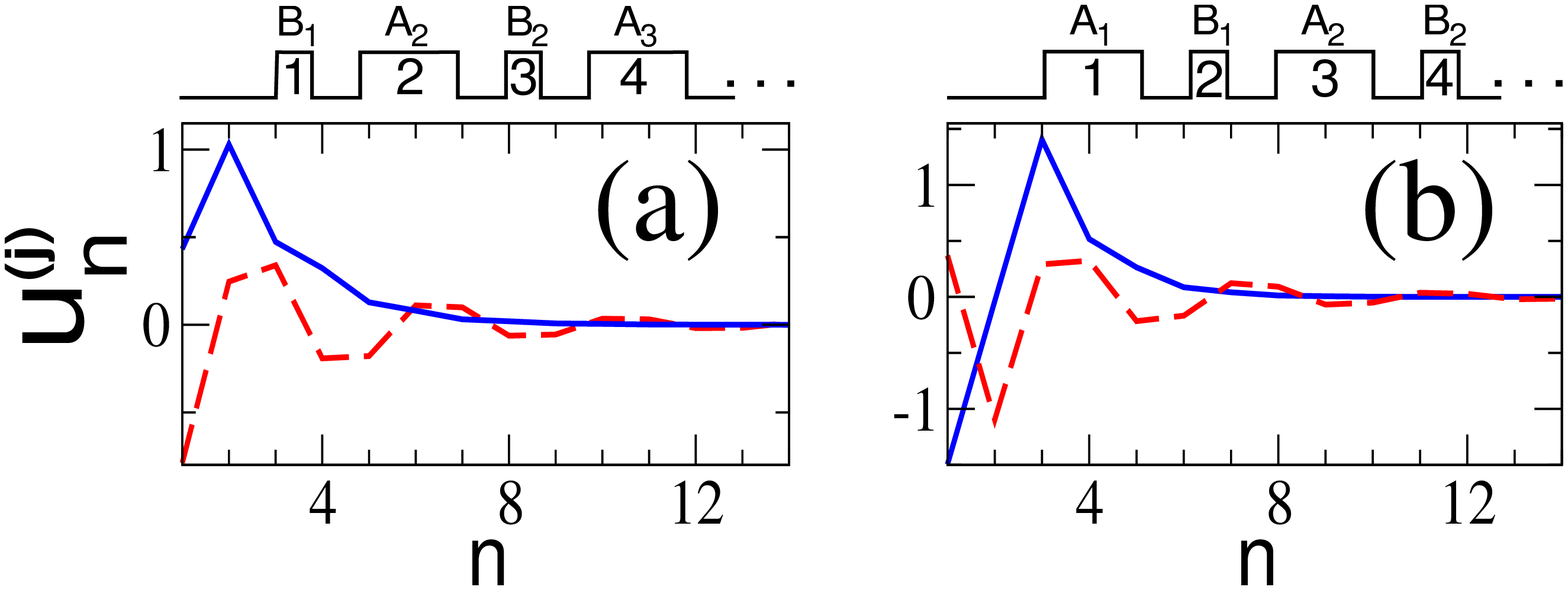}{multiGapProfiles}{ Examples of surface
multi-gap vector solitons for different structure terminations;
solid and dashed curves show the profiles of the soliton components
from the TIR and BR gaps, respectively. For the termination on the
narrow site (a), BR component propagation constant is $\beta=-0.1$
and its power is $P=0.9$. TIR component has $\beta=2.6$ and $P=1.6$.
For the termination on the wide site (b), BR component propagation constant and power are $\beta=0$ and $P=1.6$, respectively, while TIR component has
$\beta=3.0$ and $P=4.6$.}

\pict[0.75]{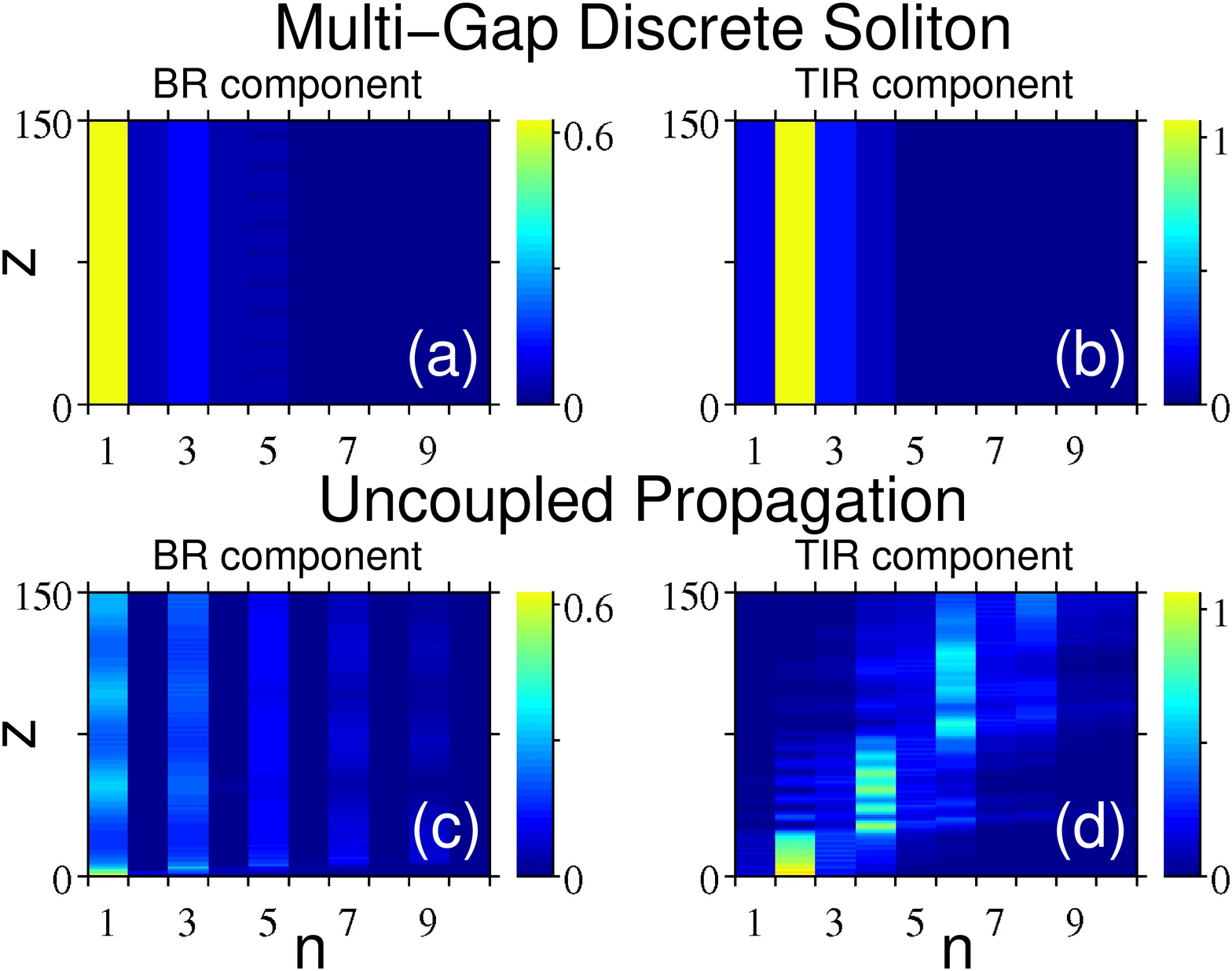}{multiGapDynamics}{ Density plots for the
intensity evolution of the field components with the profiles shown
in Fig.~\rpict{multiGapProfiles}(a): (a,b)~Stable propagation of an
initially perturbed multi-gap soliton; (c,d)~Breathing and decay of
individually propagating components. }

\section{Multi-gap surface solitons}

When the amplitude of the guided mode grows, the mode interacts with
the surface soliton waveguide creating a coupled multi-gap state in
the form of {\em a surface multi-gap vector soliton}. The
eigenvalues of the linear guided modes define the point where such
vector solitons bifurcate from their scalar
counterparts~\cite{Cohen:2003-113901:PRL, Sukhorukov:2003-113902:PRL}. In the vicinity of
the bifurcation point, the soliton symmetry and stability are
defined mostly by the large-amplitude soliton component. However, as
the power in the second component grows, the soliton properties
change dramatically, and the surface multi-gap vector solitons may
demonstrate quite complicated structure and behavior.
Figures~\rpict{multiGapProfiles}(a,b) show two examples of such
two-gap surface solitons found numerically for the two different
truncations of the binary waveguide array. In both cases, the
composite state is created by coupling the components residing in
the TIR and BR spectral gaps. Properties of the surface multi-gap
solitons can be engineered by controlling geometry and parameters of
the semi-infinite array. For example, in the case of the termination
of the binary array on the wide site, TIR component of the stable
multi-gap soliton has a twisted structure [see
Fig.~\rpict{multiGapProfiles}(b)] which reflects the fact that when
the array is truncated on the wide site, a fundamental TIR soliton can
not support any guided mode in the other gap
[Fig.~\rpict{solitonsModesWide}(c)]. We find that these multi-gap
solitons are stable for propagation over more than 500 coupling
lengths under initial perturbations on the order of 1\%
[Figs.~\rpict{multiGapDynamics}(a,b)], whereas in the absence of
mutual trapping, the individual TIR and BR components experience
breathing and decay [Figs.~\rpict{multiGapDynamics}(c,d)].

\section{Conclusion}

We have introduced a novel type of nonlinear surface waves created
by mutual trapping of several components from different spectral
gaps in semi-infinite systems with multi-gap transmission spectra.
We have studied surface modes in truncated binary waveguide
arrays and found various multi-gap surface states including the
surface soliton-induced waveguides with the guided modes from different
 spectral gaps and multi-gap composite vector solitons.

\section*{Acknowledgements}

The authors acknowledge support from the Australian Research Council
and Fondecyt (grants 1050193 and 7050173).

\end{sloppy}

\begin{thebibliography}{10}

\bibitem{Davidson:1996:SurfaceStates}
S.~G. Davidson and M. Steslicka, {\em {Basic theory of surface states}} (Oxford
  Science Publications, New York, 1996).

\bibitem{Tamm:1932-849:ZPhys}
I.~E. Tamm, ``A possible kind of electron binding on crystal surfaces,'' Z.
  Phys. {\bf 76,} 849--850 (1932).

\bibitem{Yeh:1977-423:JOS}
P. Yeh, A. Yariv, and C.~S. Hong, ``Electromagnetic propagation in periodic
  stratified media .1. General theory,'' J. Opt. Soc. Am. {\bf 67,} 423--438
  (1977).

\bibitem{Yeh:1978-104:APL}
P. Yeh, A. Yariv, and A.~Y. Cho, ``Optical surface waves in periodic layered
  media,'' Appl. Phys. Lett. {\bf 32,} 104--105 (1978).

\bibitem{Tomlinson:1980-323:OL}
W.~J. Tomlinson, ``Surface wave at a nonlinear interface,'' Opt. Lett. {\bf 5,}
  323--325 (1980).

\bibitem{Boardman:1991-73:NonlinearSurface}
A.~D. Boardman, P. Egan, F. Lederer, U. Langbein, and D. Mihalache,
  ``Third-order nonlinear electromagnetic TE and TM guided waves,''  in {\em
  Nonlinear Surface Electromagnetic Phenomena}, Vol.~29 of {\em Modern Problems
  in Condensed Matter Sciences}, H.~E. Ponath and G.I. Stegeman, eds.,
  (North-Holland, Amsterdam, 1991), \ pp.\ 73--287.

\bibitem{Mihalache:1989-229:ProgressOptics}
D. Mihalache, M. Bertolotti, and C. Sibilia, ``Nonlinear wave propagation in
  planar structures,''  in {\em Progress in Optics}, E. Wolf, ed.,
  (North-Holland, Amsterdam, 1989), Vol.~XXVII, \ pp.\ 229--313.

\bibitem{Kivshar:1998-125:PD}
Yu.~S. Kivshar, F. Zhang, and S. Takeno, ``Multistable nonlinear surface
  modes,'' Physica D {\bf 119,} 125--133 (1998).

\bibitem{Makris:2005-2466:OL}
K.~G. Makris, S. Suntsov, D.~N. Christodoulides, G.~I. Stegeman, and A. Hache,
  ``Discrete surface solitons,'' Opt. Lett. {\bf 30,} 2466--2468 (2005).

\bibitem{Suntsov:2006-63901:PRL}
S. Suntsov, K.~G. Makris, D.~N. Christodoulides, G.~I. Stegeman, A. Hache, R.
  Morandotti, H. Yang, G. Salamo, and M. Sorel, ``Observation of discrete
  surface solitons,'' Phys. Rev. Lett. {\bf 96,} 063901--4 (2006).

\bibitem{Kivshar:2003:OpticalSolitons}
Yu.~S. Kivshar and G.~P. Agrawal, {\em {Optical Solitons: From Fibers to
  Photonic Crystals}} (Academic Press, San Diego, 2003).

\bibitem{Molina:2006-discrete:OL}
M. Molina, R. Vicencio, and Yu.~S. Kivshar, ``Discrete solitons and nonlinear
  surface modes in semi-infinite waveguide arrays,'' Opt. Lett. 31 (2006), in
  press.

\bibitem{Kartashov:2006-73901:PRL}
Y.~V. Kartashov, V.~A. Vysloukh, and L. Torner, ``Surface gap solitons,'' Phys.
  Rev. Lett. {\bf 96,} 073901--4 (2006).

\bibitem{Rosberg:2006-xxx:ProcCLEO}
C.~R. Rosberg, D.~N. Neshev, W. Krolikowski, Yu.~S. Kivshar, A. Mitchell, R.~A.
  Vicencio, and M.~I. Molina, ``Observation of surface gap solitons,'' In {\em
  Conference on Lasers and Electro-Optics (CLEO), OSA Technical Digest}, \ p.\
  CMK7  (Optical Society of America, Washington DC, 2006).

\bibitem{Hudock:2005-7720:OE}
J. Hudock, S. Suntsov, D.~N. Christodoulides, and G.~I. Stegeman, ``Vector
  discrete nonlinear surface waves,'' Opt. Express {\bf 13,} 7720--7725 (2005),
  \url{http://www.opticsexpress.org/abstract.cfm?URI=OPEX-13-20-7720}.

\bibitem{Cohen:2003-113901:PRL}
O. Cohen, T. Schwartz, J.~W. Fleischer, M. Segev, and D.~N. Christodoulides,
  ``Multiband vector lattice solitons,'' Phys. Rev. Lett. {\bf 91,} 113901--4
  (2003).

\bibitem{Sukhorukov:2003-113902:PRL}
A.~A. Sukhorukov and Yu.~S. Kivshar, ``Multigap discrete vector solitons,''
  Phys. Rev. Lett. {\bf 91,} 113902--4 (2003).

\bibitem{Cohen:2005-500:NAT}
O. Cohen, G. Bartal, H. Buljan, T. Carmon, J.~W. Fleischer, M. Segev, and D.~N.
  Christodoulides, ``Observation of random-phase lattice solitons,'' Nature
  {\bf 433,} 500--503 (2005).

\bibitem{Rosberg:2005-5369:OE}
C.~R. Rosberg, B. Hanna, D.~N. Neshev, A.~A. Sukhorukov, W. Krolikowski, and
  Yu.~S. Kivshar, ``Discrete interband mutual focusing in nonlinear photonic
  lattices,'' Opt. Express {\bf 13,} 5369--5376 (2005),
  \url{http://www.opticsexpress.org/abstract.cfm?URI=OPEX-13-14-5369}.

\bibitem{Sukhorukov:2002-2112:OL}
A.~A. Sukhorukov and Yu.~S. Kivshar, ``Discrete gap solitons in modulated
  waveguide arrays,'' Opt. Lett. {\bf 27,} 2112--2114 (2002).

\bibitem{Sukhorukov:2003-2345:OL}
A.~A. Sukhorukov and Yu.~S. Kivshar, ``Generation and stability of discrete gap
  solitons,'' Opt. Lett. {\bf 28,} 2345--2347 (2003).

\bibitem{Morandotti:2004-2890:OL}
R. Morandotti, D. Mandelik, Y. Silberberg, J.~S. Aitchison, M. Sorel, D.~N.
  Christodoulides, A.~A. Sukhorukov, and Yu.~S. Kivshar, ``Observation of
  discrete gap solitons in binary waveguide arrays,'' Opt. Lett. {\bf 29,}
  2890--2892 (2004).

\bibitem{Pelinovsky:2004-36618:PRE}
D.~E. Pelinovsky, A.~A. Sukhorukov, and Yu.~S. Kivshar, ``Bifurcations and
  stability of gap solitons in periodic potentials,'' Phys. Rev. E {\bf 70,}
  036618--17 (2004).

\end{thebibliography}
\end{document}